\documentclass[journal=jacsat,manuscript=article]{achemso}

\usepackage{chemformula} 
\usepackage[T1]{fontenc} 

\author{Prashant Raj}
\author{Mishu Paul}
\author{Mythreyi R.}
\author{Balanarayan Pananghat}
\email{balanarayan@iisermohali.ac.in}
\affiliation{Department of Chemical Sciences, Indian Institute of Science Education and Research (IISER) Mohali, Knowledge City, S. A. S. Nagar, Mohali, Punjab-140306, India.}

\title[An \textsf{achemso} demo]
  {Bonding in light-induced vortices: benzene in a high-frequency circular polarized laser\footnote{with thanks to Professor Srihari Keshavamurthy (IIT-Kanpur) for the term \emph{light-induced vortices}}}

\begin{document}

\begin{tocentry}
\includegraphics[height=3.5cm,width=9cm]{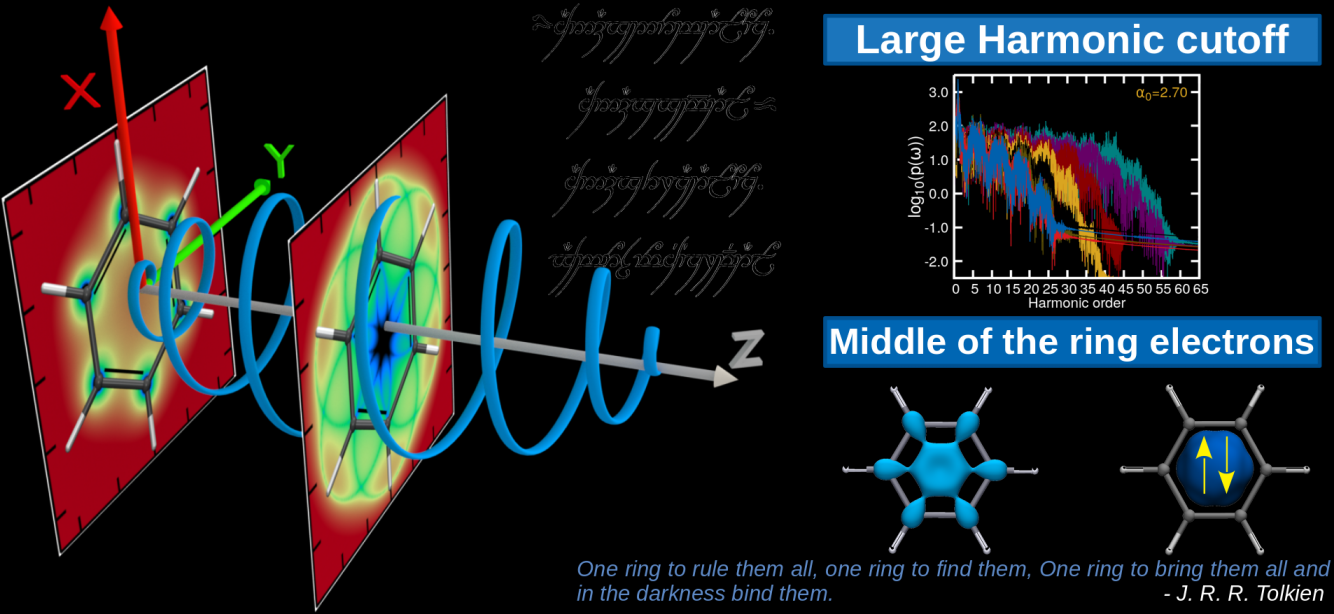}
\end{tocentry}

\newpage
\begin{abstract}
The electronic structure of benzene in the presence of a high-intensity high-frequency circularly polarized laser supports a middle-of-the-ring electron localization. Here, the laser polarization coincides with the ring plane of benzene.  The high-frequency oscillating electric field creates circular currents centered at each atom with a circle radius equal to the maximum field amplitude of the laser.  All six carbons have six such rings. For a maximum field amplitude of 1.42 \AA, which is the carbon-carbon bond distance, all six dynamic current circles intersect to create a deep vortex in the middle, which supports a bound state of a pair of electrons. Such states for benzene can be realized in experiments using a circularly polarized XUV-laser in a range of intensities $10^{16}$-$10^{17}$ W/cm$^{2}$ and frequencies $16$ eV to $22$ eV. Electronic dynamics calculations predict a minimal ionization of benzene when the rise-time of the laser pulse is sudden, indicating a possible experimental realization of these states characterized by a large cut-off in the harmonic generation spectra. This stable electronic structure of the light-dressed benzene  is doubly-aromatic due to an extra aromaticity from a D$_{6h}$ symmetric circular distortion of the $\sigma$-framework while the $\pi$-electrons, with low density in the ring-plane, are least affected.
\end{abstract}
\newpage
\section{Introduction}
\par Contrary to intuition and expectation, high-frequency laser fields with intensities comparable to internal electric fields of atoms or molecules create long-lived quasi-bound states, which are almost stable to decay, dissociation or ionization.\cite{Pont1988,Eberly1993,Aubanel1993} Chemistry in these strong high-frequency fields is not far from reality with the maximum achievable laser intensity now being recorded at $10^{21}$ W/cm$^{2}$ consisting of photons that are almost in the relativistic regime. The existence of quasi-bound states (metastable, but slow-ionizing with a long lifetime) have been a theoretical prediction, with an experimental realization in the case of noble gas atoms such as neon, helium and argon.\cite{Boer1993,Smirnova2003,Eichmann2009,Eichmann2013,Matthews2018} Over the decades, many theoretical studies have looked at such exotic states in the case of diatomic molecules.\cite{Yasuike2004,Wei2008,Wei2018} Several interesting effects such as conformational changes for small molecules, linear Stark shifts of laser-dressed atoms etc. have been theoretically proposed.\cite{Balanarayan2012,Kumar2019} A  question that always springs up for a chemist is what happens with polyatomic molecules? How would their light-dressed stable states look like?\cite{Protopapas1997}
\par With several theoretical proposals of driving electronic currents in molecules such as porphyrin\cite{Barth2006, Nobusada2007,Kanno2018}, and some latest path-breaking experiments on graphene,\cite{Bunch2008,Jiang2009,Miao2013,Hu2014,Kroes2017,Heide2018,Poltavsky2018} a simple molecule of thought for a chemist would be benzene. Given its $D_{6h}$ geometry and aromaticity, circularly  polarized lasers (CPLs) would be an apt choice to understand its electronic dynamics in a strong high-frequency oscillating field. For porphyrin, the proposal was to induce a unidirectional electron flow through its structure with a CPL.\cite{Barth2006} A quantum control of electronic fluxes charge migration in the attosecond regime was demonstrated for a benzene molecule, using a combination of linearly and circularly polarized lasers. The laser preparation was found to impart electronic phases to the model molecule, affecting the charge migration mechanism.\cite{Jia2017}
\par Since attosecond, strong pulses of circularly polarized light are the call of the day, strong-field effects in benzene invoke ample curiosity, for both experiments and theory.\cite{Liu2015,Liu2019,Jia2017,Winney2017} The latest experiments on graphene have come up with new topological states created by laser pulses.\cite{McIver2019} Even though the electronic structure of benzene is different from that of graphene, as the latter comprises a band structure that lacks a band gap, benzene can be considered a \emph{beginner's choice} apropos model for graphene.  A study of benzene in CPL fields of $10^{14}$-$10^{15}$ W cm$^{-2}$ by Baer et al. shows a preservation of the D$_{6h}$ symmetry, interpreted from a higher harmonic generation spectra.\cite{Baer2003} The CPL-induced dynamics and laser-dressed electronic structure of benzene need to be understood in perspectives of contrast as well as similarities with graphene.
\par In a circularly polarized laser, the time-dependent electric field with $x$- and $y$-polarization in Cartesian directions (and a propagation direction of $z$) will be,
 \begin{equation}
 \vec{\epsilon}(t) = f(t)\epsilon_{0}\left[\cos(\omega t)\hat{e}_{x} +\sin(\omega t)\hat{e}_{y} \right],
 \label{eq:eq01}
 \end{equation}
 in a pulse envelope, $f\left(t\right)$, and peak electric field strength, $\epsilon_{0}$, with,  $\omega$, as the frequency of the laser. For a continuous wave (CW) laser we set $f\left(t\right)=1$, with an instantaneous rise-time. For benzene, in this work, a CW-laser form (as a CPL) is chosen initially instead of a pulse, to understand its light-dressed state and possible light-dressed effects.
The Time-Dependent Schr\"odinger Equation (TDSE) for such an $n$-electron, $M$-atom molecule is (in atomic units, a.u.):  
 \begin{eqnarray}
i\frac{\partial}{\partial t}\Psi\left(\{\vec{r}_{j}\},t\right)=\left[\frac{1}{2}\sum_{j}^{n} \left[\hat{p}_{j} + \hat{\vec{A}}(t) \right]^{2} - \sum_{A}^{M}\sum_{j}^{n}\dfrac{Z_{A}}{\left| \vec{r}_{j} - \vec{R}_{A} \right|} + \sum_{j>k}^{n}\frac{1}{\left|\vec{r}_{j}-\vec{r}_{k}\right|}\right]\Psi\left(\{\vec{r}_{j}\},t\right),
\label{eq:eq02}
\end{eqnarray}
where, $\vec{A}(t)$ is a vector potential with $\vec{\epsilon}\left(t\right)=-\dfrac{d\vec{A}(t)}{dt}$ in a.u. Moving on to the Kramers-Henneberger (KH) frame of reference, which  is an oscillating frame of reference, is achieved via a time-dependent unitary transformation\cite{Kramers1956, Henneberger1968}:
\begin{eqnarray}
\hat{\Omega}=\exp{\left[i\int^{t}_{-\infty}{\left(\vec{A}\left(\tau\right)\cdot\sum_{j=1}^{n}\vec{p}_{j}+\frac{1}{2}\left|\vec{A}\left(\tau\right)\right|^{2}\right)d\tau}\right]}. 
\label{eq:eq03}
\end{eqnarray}
In the KH frame, the wave function $\Phi\left(\{\vec{r}_{j}\},t\right)$ is $\hat{\Omega} \Psi\left(\{\vec{r}_{i}\},t\right)$, a unitarily transformed quantity. Here, $\hat{\Omega}=\hat{\Omega}_{1}\hat{\Omega}_{2}$, with $\hat{\Omega}_{1}=\exp{\left[i\sum_{j=1}^{n}{\int^{t}_{-\infty}{\vec{A}\left(\tau\right)\cdot\vec{p}_{j}d\tau}}\right]}$. This gives a TDSE of the form:
\begin{eqnarray}
i\frac{\partial}{\partial t}\Phi\left(\{\vec{r}_{j}\},t\right)=\left[ \frac{1}{2}\sum_{j}^{n}\hat{p}^{2}_{j} +\sum_{j=1}^{n}{V_{C}\left[\vec{r}_{j}-\vec{\alpha}\left(t\right)\right]}
 + \sum_{j>k}^{n}\frac{1}{\left|\vec{r}_{j}-\vec{r}_{k}\right|}\right]\Phi\left(\{\vec{r}_{j}\},t\right).
\label{eq:eq04}
\end{eqnarray}
The effect of $\hat{\Omega}_{2}$ is to generate a purely time-dependent phase factor while $\hat{\Omega}_{1}$ acts as a time-dependent displacement operator, modifying the nuclear-electron attraction potential to the expression:
\begin{eqnarray}
V_{C}\left[\vec{r}-\vec{\alpha}\left(t\right)\right]=-\sum_{A}^{M}\dfrac{Z_{A}}{\left| \vec{r} - \alpha_{0}\hat{e}_{x}\cos(\omega t) - \alpha_{0}\hat{e}_{y}\sin(\omega t)- \vec{R}_{A} \right|}. 
\label{eq:eq05}
\end{eqnarray}
The inter-electron repulsion remains the same. Explicitly, $\vec{r}_{j}\rightarrow \vec{r}_{j}-\vec{\alpha}\left(t\right)$ and $\vec{r}_{k}\rightarrow \vec{r}_{k}-\vec{\alpha}\left(t\right)$, and consequently in $\left|\vec{r}_{j}-\vec{r}_{k}\right|$, the time-dependent shift cancels. In simple intuitive terms, the inter-electronic (particle) distance will remain the same, on a change in the frame of reference. The factor $\alpha_{0}$ has units of distance and is the maximum field amplitude of the laser, in a.u.:
\begin{eqnarray}
\alpha_{0}=\frac{\epsilon_{0}}{\omega^{2}}.
\label{eq:eq06}
\end{eqnarray}
 \par Thus, in this oscillating frame of reference the binding Coulomb potential is time dependent. For an isolated single atom, the binding potential has a form where the Coulombic well (singularity) traverses in a circular path with the radius of the circle being $\alpha_{0}$, the maximum field amplitude of the CPL. This is the \emph{oscillating} frame of reference from the point of view of the electron. For such a spatially circular and temporally periodic traversal of a real potential, the harmonics can be understood via a time-dependent Fourier expansion \cite{Smirnova2003} symbolized as:
\begin{eqnarray}
V_{C}\left[\vec{r}-\vec{\alpha}\left(t\right)\right]=V_{0}^{KH}\left(\vec{r}\right)+\sum^{+\infty}_{n=1}{V^{KH}_{n}\left(\vec{r}\right)}\cos\left(n\omega t\right).
\label{eq:eq07}
\end{eqnarray}
In the break-up of the sum, the individual Fourier components (harmonics) have the form:
\begin{eqnarray}
V^{KH}_{n}\left(\vec{r}\right)=\frac{1}{T}\int^{T}_{0}{V_{C}\left[\vec{r}-\alpha_{0}\left(\hat{e}_{x}\cos{\omega t}+\hat{e}_{y}\sin{\omega t}\right)\right]\cos{(n\omega t)}dt}.
\label{eq:eq08}
\end{eqnarray}
over $T=\frac{2\pi}{\omega}$, the time period. 
\par For a high-frequency CPL, on a time averaging (over the faster, high-frequency $n>0$ components), the temporally oscillating terms, $V^{KH}_{n}\left(\vec{r}\right)$ with $n>0$ would contribute less. Hence, the potential term, $V_{0}^{KH}\left(\vec{r}\right)$, will now be the effective binding potential instead of the field-free Coulomb potential. The zeroeth order term $V^{KH}_{0}\left(\vec{r}\right)$ is time independent (the first term in a Fourier expansion is a simple time average) and the \emph{dominant} one, which gives a time-independent Schr\"odinger equation in the effective potential. \cite{Pont1988,Gavrila2002} It was shown by solving Equation \ref{eq:eq04} using high frequency Floquet theory (HFFT) in successive lowest order of $\omega^{-1}$, that Equation \ref{eq:eq04}  reduces to a time-independent Schr\"odinger equation with the potential given by $ V_{0}^{KH}$. 
In the limit of a high-frequency, an adiabatic Floquet theory ensures that the time-dependence is rendered inconsequential, as the higher order terms in the KH potential are negligible. Thus, for short laser pulses, ionization is suppressed due to a laser-induced resonance state.\cite{Barash1999} Using a laser intensity as high as 10$^{16}$ W/cm$^{-2}$ and frequency $806$ nm, experiments have demonstrated that the electron scattering is "switched off" and ionization and molecular fragmentation are suppressed in these regimes.\cite{Rajgara2003} Currently, lasers with higher intensities, frequency of 26.5 eV and pulse that lasts upto 10 ns are available.\cite{Heinbuch2005}  
The high frequency conditions also ensure that the ionization is minimized and the bound states of the time-independent Schr\"odinger equation in the effective potential determine the chemistry.\cite{Faisal2000}
\begin{figure}[t]
\includegraphics[width=0.9\linewidth]{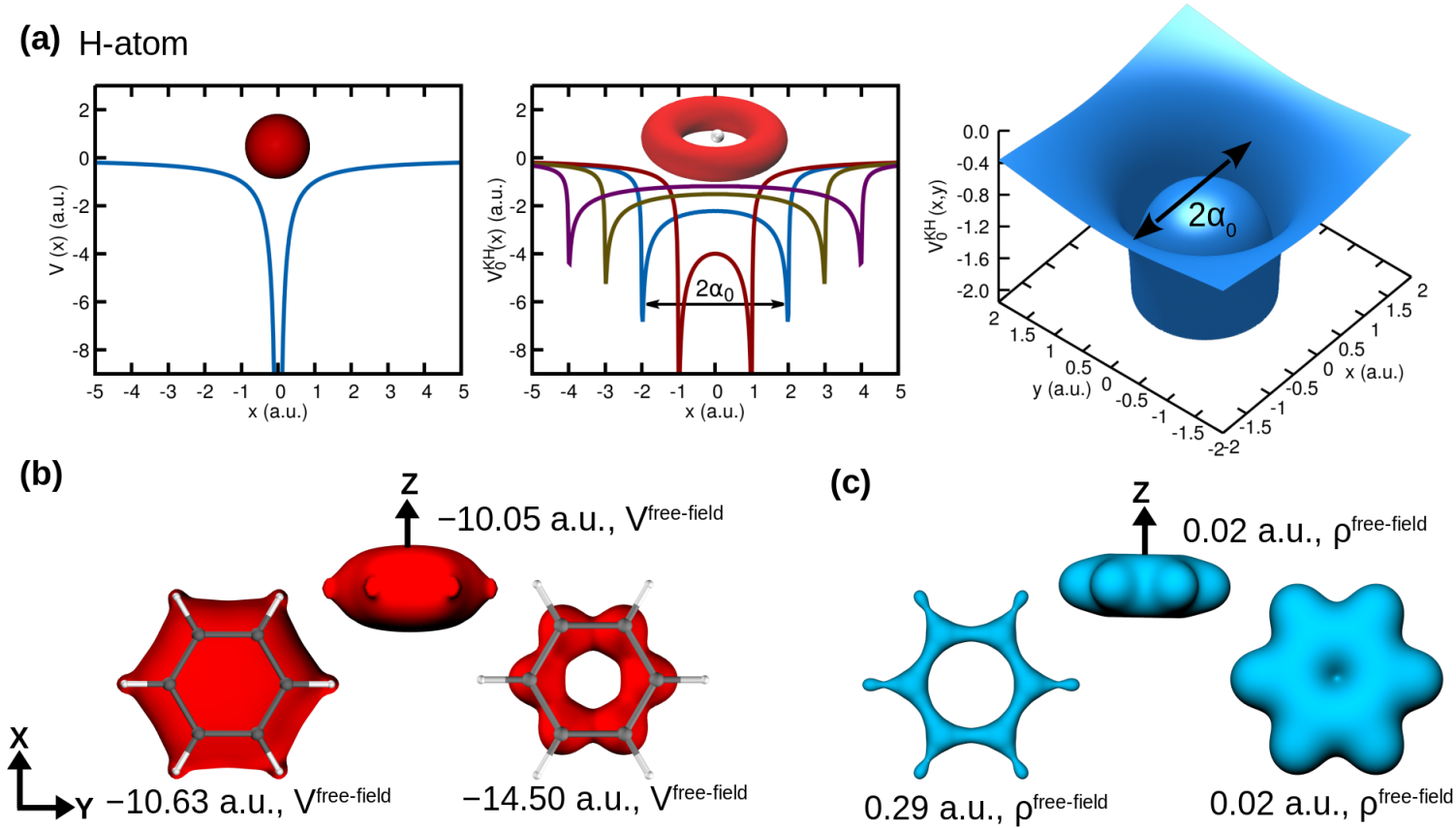}
\caption{Motivation and Introduction : \textbf{(a)} 1-D plots for a without-field and with-CPL isolated single atom. The spherical Coulomb potential loses its singularity upon switching on the laser and traverses a circular path with the radius of the circle being $\alpha_{0}$, the maximum field amplitude of the CPL, depicted as a 'split' in the potential, revolution of which would form the figure adjacent in 2-D. A 3-D plot for an effective KH-potential. 
Isosurface plots for field-free \textbf{(b)} bare-nuclear potential (V$^{field-free}$) and \textbf{(c)} Electron Density ($\rho$). }
\label{fig:fig1}
\end{figure}
\par In this zeroeth order time-independent effective potential, the Coulombic binding potential transforms from spherical to toroidal. This is portrayed in Figure~\ref{fig:fig1}\textbf{(a)} where the usual spherical Coulomb potential for a single, isolated atom and the corresponding high-frequency laser-dressed potential (see Figure ~\ref{fig:fig1}\textbf{(a)}, second and third plots) have been depicted. The isosurfaces for the Coulomb potential assume a \emph{doughnut-like} toroidal shape with a saddle point at the atomic position instead of the Coulombic singularity. Thus, when the laser intensity is high enough (high electric field strength), the electric field of the CPL induces a circular motion of the electrons in the plane of polarization, the circle being centered at the atom. The radius of this torus is equal to $\alpha_{0}$ given in Equation~\ref{eq:eq06},which in turn is proportional to the electric field strength, $\epsilon_{0}$, for a fixed $\omega$, frequency.
\begin{figure}[t]
\includegraphics[width=\linewidth]{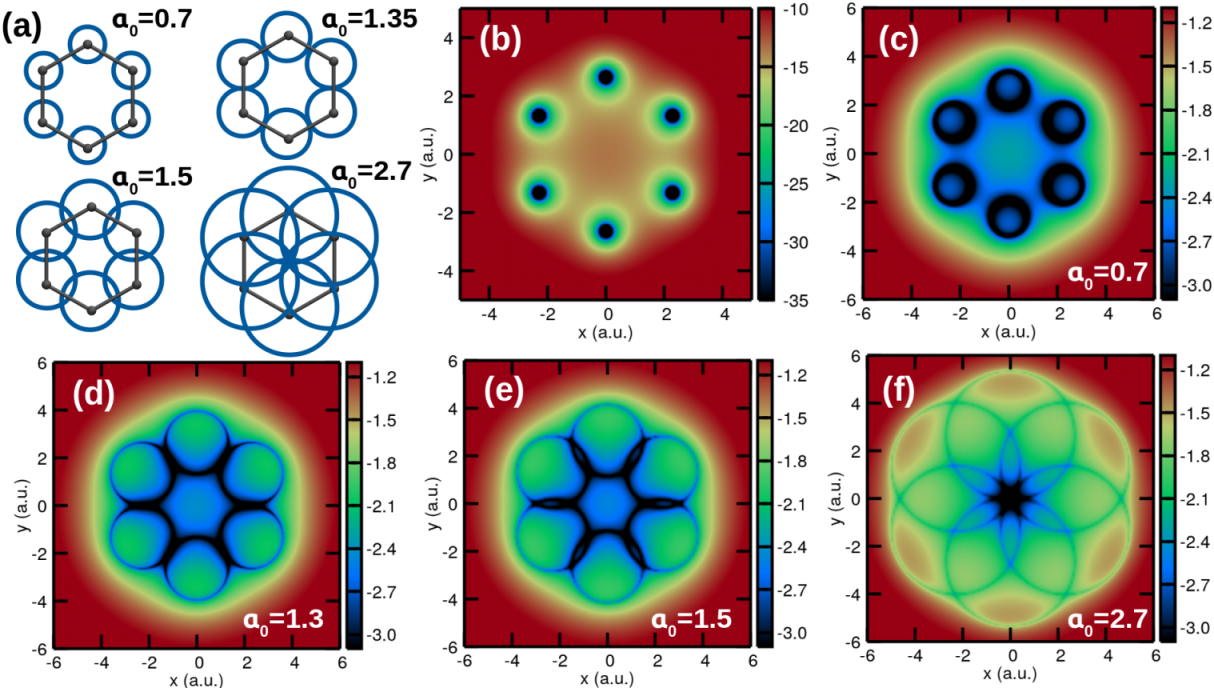}
\caption{\textbf{(a)} A schematic representation of circles centered at vertices of a hexagon, intersecting at varying $\alpha_{0}$, given in the panels \textbf{(b)} to \textbf{(f)}. At $\alpha_{0}= 2.7$ a.u. the circle radius is just enough to form an optimum overlap at the hexagon center, giving the deepest point in the laser dressed potential to be the middle-of-the ring, which is expected to localize electrons.}
\label{fig:fig2}
\end{figure}
\par In our work, the effect of a CPL on a benzene molecule has been examined with the laser polarization taken along $xy$-plane which is the molecular plane. The field-free (without laser) Coulombic potentials and electron densities of benzene molecule are presented in Figure~\ref{fig:fig1}\textbf{(b)} and \textbf{(c)} where the field-free Coulomb potential for benzene has an atom-centered structure with singularities at the carbon centers. 
For the one-electron density\cite{Morrell1975,Parr1989,Bader1990} there are maxima at the nuclear positions and saddle points at bond-centers, bringing out the structural signatures of field-free benzene.
 \par The gist of the questions addressed in this work (and a hint of the results), is pictorially represented in Figure~\ref{fig:fig2} showing a two-dimensional depiction of the effective potential $V^{KH}_{0}\left(\vec{r}\right)$ for benzene. If a single atom is toroidal or circular in the plane of laser polarization, then what are the effects, unusual or  interesting, which one could have for benzene, a molecule  where each of the carbon atoms are arranged on the vertices of a regular hexagon? Figure~\ref{fig:fig2}\textbf{(a)} gives a schematic portrayal of an intensity variation of the CPL, which effectively amounts to an increasing $\alpha_{0}$, which in turn is the radius of each of the circles centered at each carbon atom. When $\alpha_{0}<\frac{r_{CC}}{2}$, the electrons are expected to simply move in small circles centered at each carbon atom. At $\alpha_{0}=\frac{r_{CC}}{2}$, the circles just touch each other, leading to deeper potentials at the middle of the bonds. For $\alpha_{0}=r_{CC}$, all of the six circles intersect right in the middle of the regular hexagon, giving a deep light-induced vortex at the ring-center. The nature of this potential suggests a very interesting electronic structure for benzene in the presence of a strong CPL, with the possibility of electrons being seated at this ring-center.
\par The nature  of these unexpected stable states achieved in the presence of the laser is described in the subsequent sections, using a comprehensive analysis of molecular orbitals. It is shown that a pair of electrons populates a bound state of an $s$-type character in the middle of the ring, as a result of a laser-induced hybridization. If the electrons' propensity for staying towards the middle of the ring is high in the presence of the laser, then this has consequences for proton stabilization at the ring-center. Interestingly, the $\pi$-cloud of benzene is unaffected by the laser polarization in the $xy$-plane, since the $\pi$-density is negligible in the plane of the ring. This ensures that the aromatic character of benzene is preserved. However, the circular distortions in the $\sigma$-framework lead to a delocalization and  yield a new aromatic character, imparting a double aromaticity to the laser-dressed benzene.\cite{Wodrich2007,Furukawa2018} This is analyzed in terms of electron localization functions and electrostatic potentials. Finally, our work gives an experimental prescription in terms of laser parameters from calculations of electron dynamics, in the lab frame and explains how these states can be realized and stabilized by a laser pulse. Signatures of harmonic generation spectra from these peculiar electronic states are characteristized by a high cut-off in the harmonics because of the high-frequency stabilization.\cite{Alon1998,Baer2003,Wardlow2016,Bandrauk2016}
\section*{Results and Discussion}
\par
\begin{figure}[t]
\includegraphics[width=\linewidth]{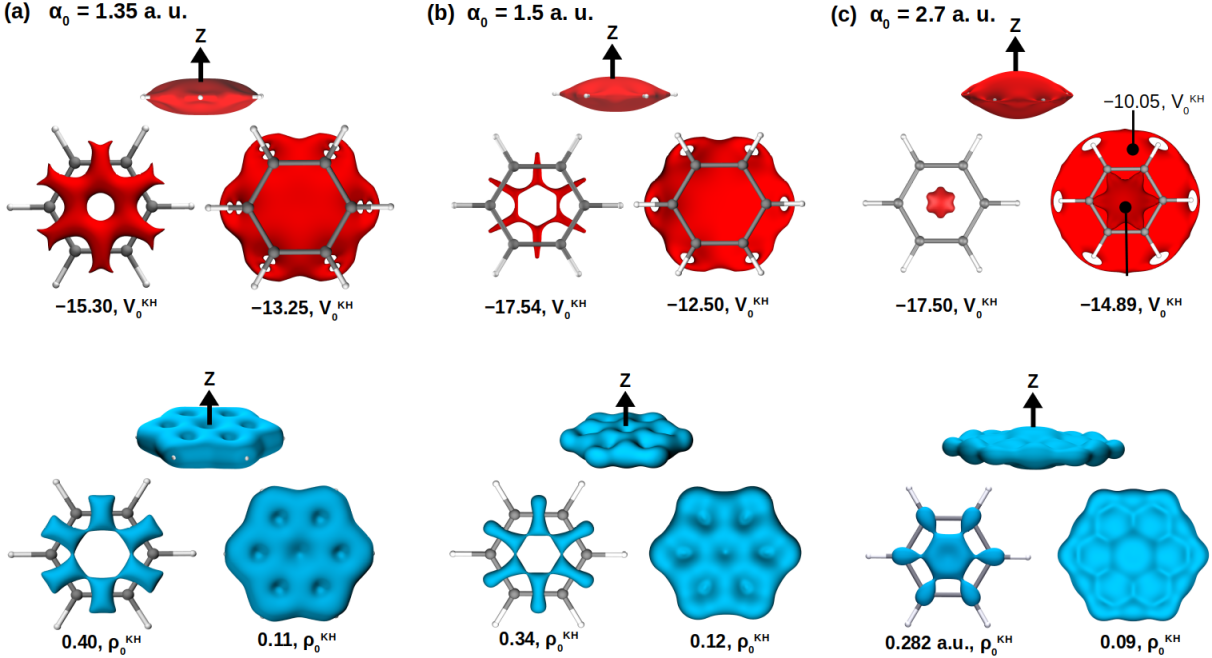}
\caption{Isosurface plots for the time-averaged zeroth-order KH potentials (Red) and Electron densities (Blue) for varying quiver distances. The potential and densities are localized at the bond centers for $\alpha_{0}$ = 1.35 a.u., and 1.50 a.u. and localized at the ring-center for $\alpha_{0}$ = 2.7 a.u.   }
\label{fig:fig3}
\end{figure}

 The first set of results are striking, and evident on an inspection of single-particle charge densities, $\rho_{KH}\left(\vec{r}\right)$ for different values of $\alpha_{0}=1.35, 1.5$ and $2.7$ a.u. These have been calculated by solving the multi-electron time-independent Schr\"odinger equation in the effective potential, $V^{KH}_{0}\left(\vec{r}\right)$, given in Equation~\ref{eq:eq06}. The methodology adopted for this, together with the basis set used have been discussed in the section on Computational Methodology. Effective potentials plotted in three-dimensions (3-D) in terms of isosurfaces are given in Figure~\ref{fig:fig3}\textbf{(a)}, \textbf{(b)} and \textbf{(c)} in the upper panel, together with different orientations of the charge densities in the lower panel. The electron densities in the effective potential give a notion of structure in the presence of the CPL. 
\par In contrast to the field-free electron density given in Figure~\ref{fig:fig1}\textbf{(c)}, the laser-dressed densities ($\rho_{0}^{KH}$) in Figure~\ref{fig:fig3} do not possess nuclei-centered maxima, neither do they satisfy the Kato's cusp condition at nuclear positions. It is obvious why this is so, from the nature of the effective potentials. Once the toroidal densities start overlapping at $\alpha_{0}\geq \frac{r_{CC}}{2}$, where $r_{CC}$ denotes the carbon-carbon distance, maxima appear in the middle of the C-C bonds, indicating a favorable bonding behavior. The hydrogens are smeared out and do not contribute much to the laser-dressed densities. Hence, instead of a nuclei-centric structure, a bond-centric structure is evident in terms of electron densities, when $\alpha_{0}=1.35$ a.u. and $\alpha_{0}=1.5$ a.u.
\par On further increasing $\alpha_{0}$, the density gets pushed towards the middle of the ring and outwards from the regular hexagon framework. This is achieved by the increasing radius of the circular currents induced by an increasing intensity of the CPL. When $\alpha_{0}=r_{CC}$, the next topological change occurs, which is visualized in terms of lower-valued densities at the atomic positions as well as beyond the inter-nuclear axis. This is because the nuclei-centric tori now intersect, or have common grounds/volume at nuclear positions, as can be seen in Figure~\ref{fig:fig2}\textbf{(f)}. A sideways horizontal view of the charge density profile given in Figure~\ref{fig:fig3}\textbf{(c)}(lower panel) now indicates a bulge in the ring-center, which is evident and in contrast to the other horizontal profiles given in the lower panels of Figure~\ref{fig:fig2}\textbf{(a)} and \textbf{(b)}. Another feature to be noted is the appearance of \emph{bonding} maxima in the immediate periphery of the ring, which is attributed to a double aromaticity, discussed later in the manuscript as a property evident from the electronic structure. A numerical integration of the densities inside the ring revealed an increase in the number of electrons, moving away from the nuclear framework to the middle of the ring. A further evidence for this will be discussed in terms of a molecular orbital analysis in the KH oscillating framework, for the time-independent effective potential. 
\par To summarize the results from the charge densities, there can be three different topological structures arising as a function of an increasing $\alpha_{0}$: (i) when $\alpha_{0}<\frac{r_{CC}}{2}$, the overlap between the tori of electron densities is the least and the density is yet to lose its nuclei-centric structure, (ii) when $\alpha_{0}\geq \frac{r_{CC}}{2}$, when the overlap is substantial between the  and then diminishing, pushing densities towards the middle of the ring and then outwards and finally (iii) when $\alpha_{0}=r_{CC}$, which corresponds to an effective potential  producing a  topological phase of maximum density at the ring-center.
\par The laser-induced \emph{structural effects}, as seen through the electron densities, can be rationalized in terms of the effective potentials given in Figure~\ref{fig:fig2}, and conceptually imagined in terms of a simple picture of six circles centered at the vertices of a regular hexagon and a count of their intersections and tangential points of contact. What happens when $\alpha_{0}>r_{CC}$. Nothing much, since a further increase in $\alpha_{0}$ makes the potential shallower. When $\alpha_{0}>r_{CC}$, the deep vortex in the middle disappears. There would always be exactly two intersections for each pair of circles and the total number of intersections therefore would be $2\times\left.^{6}C_{2}\right (=30)$, shallow points. The electron density would just keep moving away from the geometry with shallow maxima at the intersections, which would amount to undulations in an almost homogeneous electron density.
\begin{figure}[t]
\includegraphics[width=0.9\linewidth]{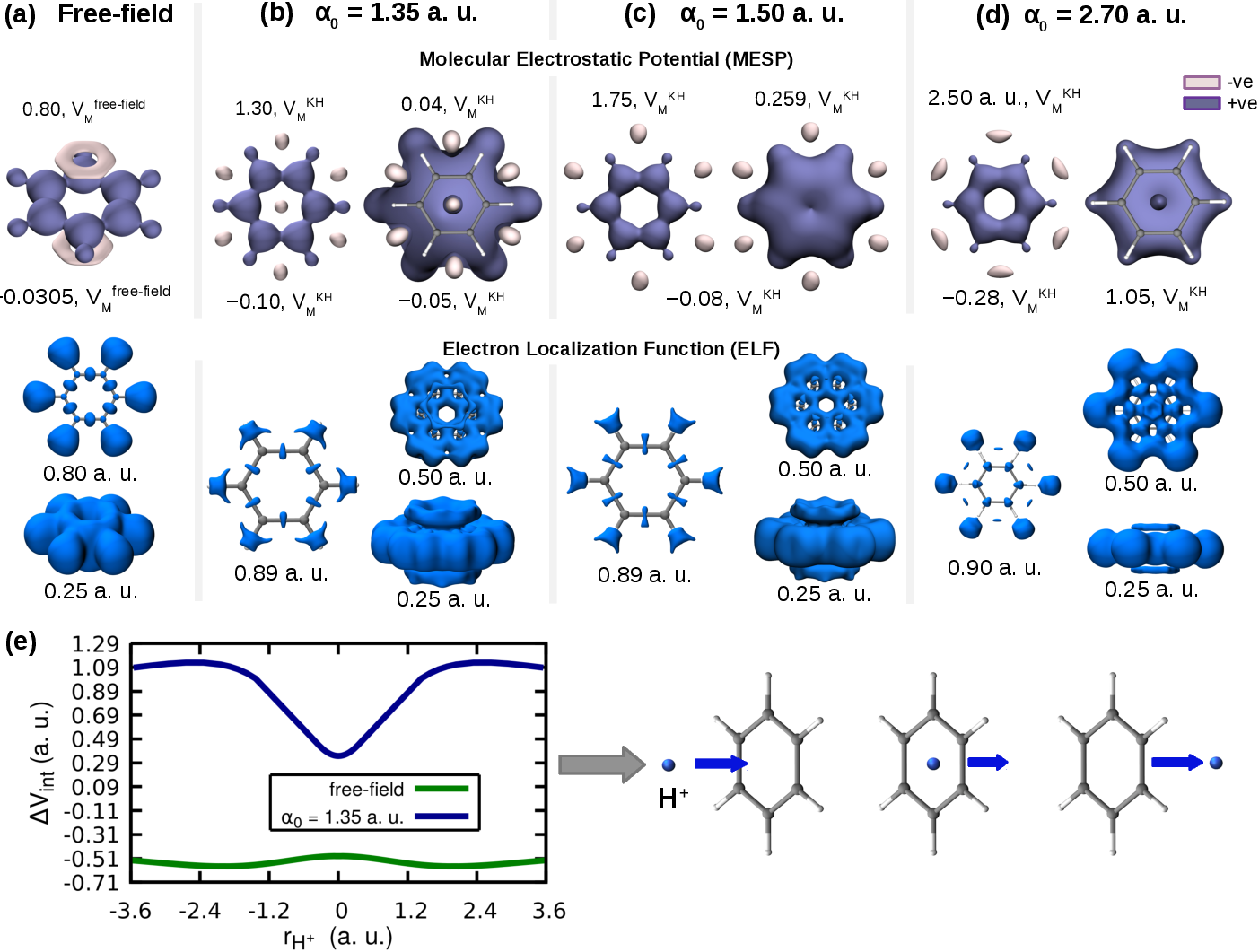}
\caption{Isosurface Plots of Molecular Electrostatic Potential (MESP ; Purple and Light-pink color scheme) and Electron Localization Function (ELF ; Blue color scheme) for benzene with CPL-field of varying quiver distances. Lower panel : Interaction Energy plots for with-field and field-free $C_{6}H_{6}$-$H^{+}$, $\Delta V_{int}$ is a minimum at the ring-center, in contrast with the barrier seen in the field-free case.}
\label{fig:fig4}
\end{figure}
\par With the laser-dressed electron density yielding structural information, it is useful to understand the \emph{reactivity} of these unusual densities. The response of the laser-dressed density to external perturbations is quantified and understood here, through properties derived from the electron density \emph{viz.}, the electrostatic potential, electron localization function and energy of interaction with a proton. The electrostatic potential\cite{Gadre1992,Gadre2000} here is defined as:
\begin{eqnarray}
V\left(\vec{r}\right)=\sum_{A}{\frac{Z_{A}}{\left|\vec{r}-\vec{R}_{A}\right|•}}-\int{\frac{\rho_{0}^{KH}\left(\vec{r}^{\hspace{0.1cm}\prime}\right)}{\left|\vec{r}-\vec{r}^{\hspace{0.1cm}\prime}\right|}d^{3}r^{\prime}}. \nonumber
\end{eqnarray}
The negative regions of the electrostatic potential of field-free densities indicate regions susceptible to electrophilic attack, with negative-valued isosurfaces enclosing negative-valued minima for lone pairs, $\pi$-clouds and aromatic $\pi$-clouds. This is seen from Figure~\ref{fig:fig4} \textbf{(a)} where the electrostatic potential for field-free benzene  shows the presence of the aromatic $\pi$-cloud, with the light-pink isosurface enclosing six negative-valued minima arranged in a hexagon, on either side of the ring. The effect of the CPL with increasing $\alpha_{0}$ shows the appearance of new negative regions at bond peripherals outside the ring and a negative-valued minimum inside the ring (Figure~\ref{fig:fig4} \textbf{(b)}). This is reminiscent of electron density getting pushed towards the middle of the ring and outwards. The outward negative-valued minima signify the doubly aromatic character developed in the system by the circularly polarized laser and the middle-of-the ring negative-valued minimum, indicates that a bare proton could bind to this site of a localized electron pair. It is also seen that in the presence of the CPL the negative regions of the electrostatic potential have become more negative in comparison with the field-free electrostatic potential. Bench-stable double aromatic compounds have been synthesized and isolated,\cite{Furukawa2018} where in addition to the $\pi$-aromaticity, a $\sigma$-aromaticity is realized in a hexaheteroatom-substituted benzene dication, and the $\sigma$ electrons follow Huckel's $4n+2$ rule, being circularly delocalized in the plane of the benzene ring. With the laser-parameters used in this work, a neutral unsubstituted benzene per se is shown to contain double-aromaticity.
\par  The interaction energies of a proton with the with-field benzene ($\Delta V_{int}$) were plotted as a function of the proton distance from the ring-center. The energy is a minimum at the ring-center for the laser-dressed system, as against the energy in the field-free case where the proton sees a barrier (Figure~\ref{fig:fig4} \textbf{(e)}). This is expected since the the MESP is a minimum (a negative and positive minimum for $\alpha_{0}$ values 1.35 a.u. and 2.7 a.u., respectively) at the ring-center, which reveals the nucleophilicity of the with-field benzene, a requisite for electrophilic aromatic substitution reactions.
\par Further information is afforded by an analysis of the electron localization function (ELF) for the laser dressed densities. For a single determinant wavefunction built from Hartree Fock orbitals $\psi_{i}$, the ELF on a 3-D grid is given as\cite{Silvi1994,Savin1996},$
ELF = \frac{1} {1+(\frac{D}{D_{h}})^{2}}$, where, $
D = \dfrac{1}{2} \sum\limits_{i} \lvert \nabla \psi_{i}\rvert ^{2} - \dfrac{1}{8} \frac{\lvert \nabla \rho\rvert^{2}}{\rho} \hspace{0.2cm}; \hspace{0.3cm} D_{h} = \dfrac{3}{10} (3\pi^{2})^{\frac{5}{3}} \rho^{\frac{5}{3}}$. The plots in the middle panel of Figure~\ref{fig:fig4} show the isosurfaces for ELF and the maxima in ELF are in agreement with the proposal that there is a middle-of-the-ring electron localization. A clear signature of aromaticity is seen for $\alpha_{0}=1.5$ a.u. and $\alpha_{0}=2.7$ a.u. Outer ring maxima correspond to the doubly aromatic character that is developed. The reason for the doubly aromatic character is rationalized  with respect to the molecular orbitals (MOs) and energy levels with and without the field.
\begin{figure}[t]
\includegraphics[width=0.95\linewidth]{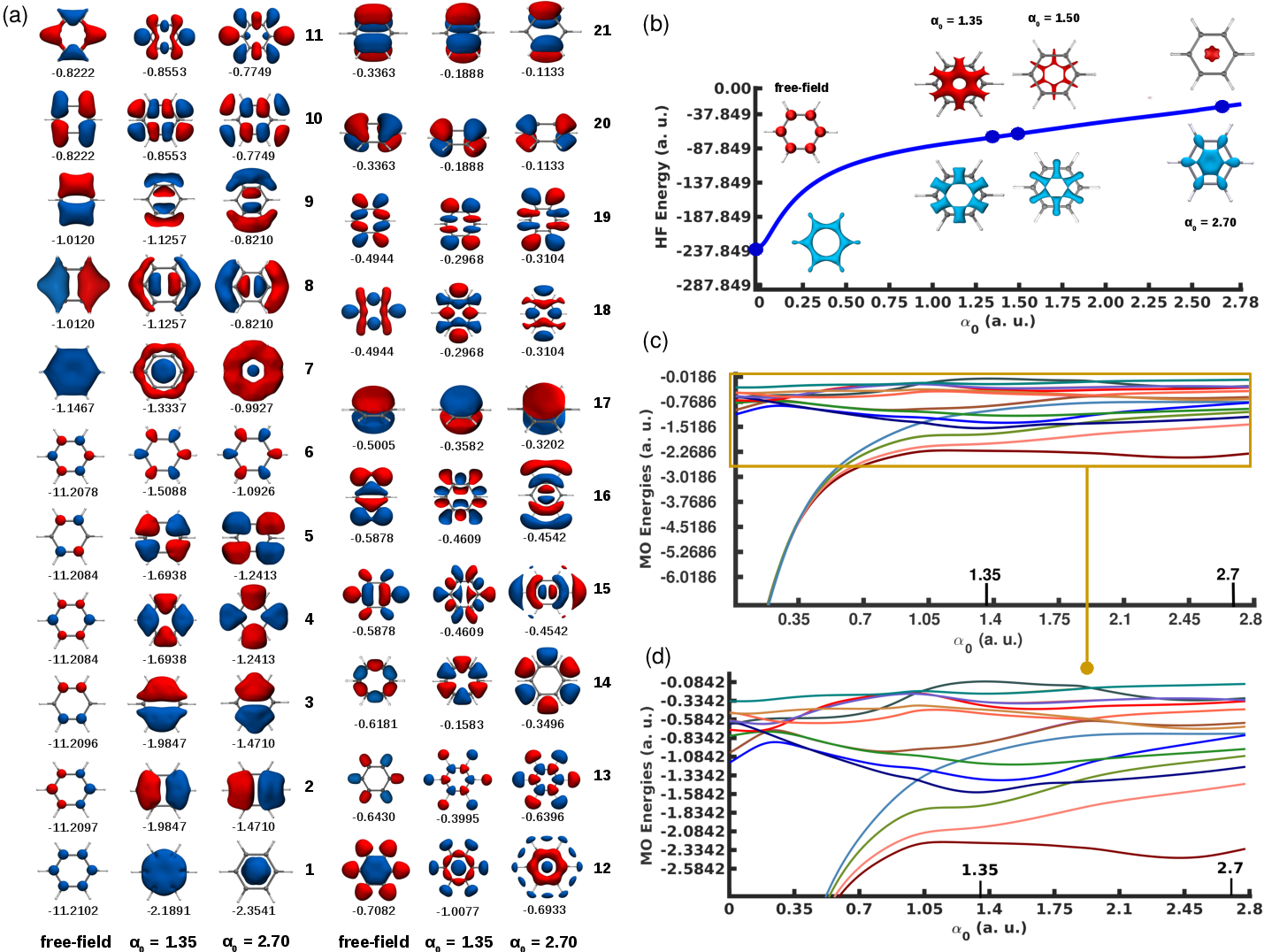}
\caption{A molecular orbital analysis for field-free and with-field benzene. (a) MOs are plotted with increasing energy. MOs 1-6 can be treated as core $\sigma$-orbitals which do not gain a node when the laser is switched on. MOs 17, 20, 21 are $\pi$-orbitals unaffected by the laser. The remaining MOs seem to have undergone the most change in the presence of a laser, with an increase in the number of nodes by one (except for 14 and 19) (b) Hartree-fock Energy plot as a function of quiver distance of the CPL. (C) MO energies as a function of the quiver distance.}
\label{fig:fig5}
\end{figure}
\par
\begin{figure}[t]
\includegraphics[width=0.8\linewidth]{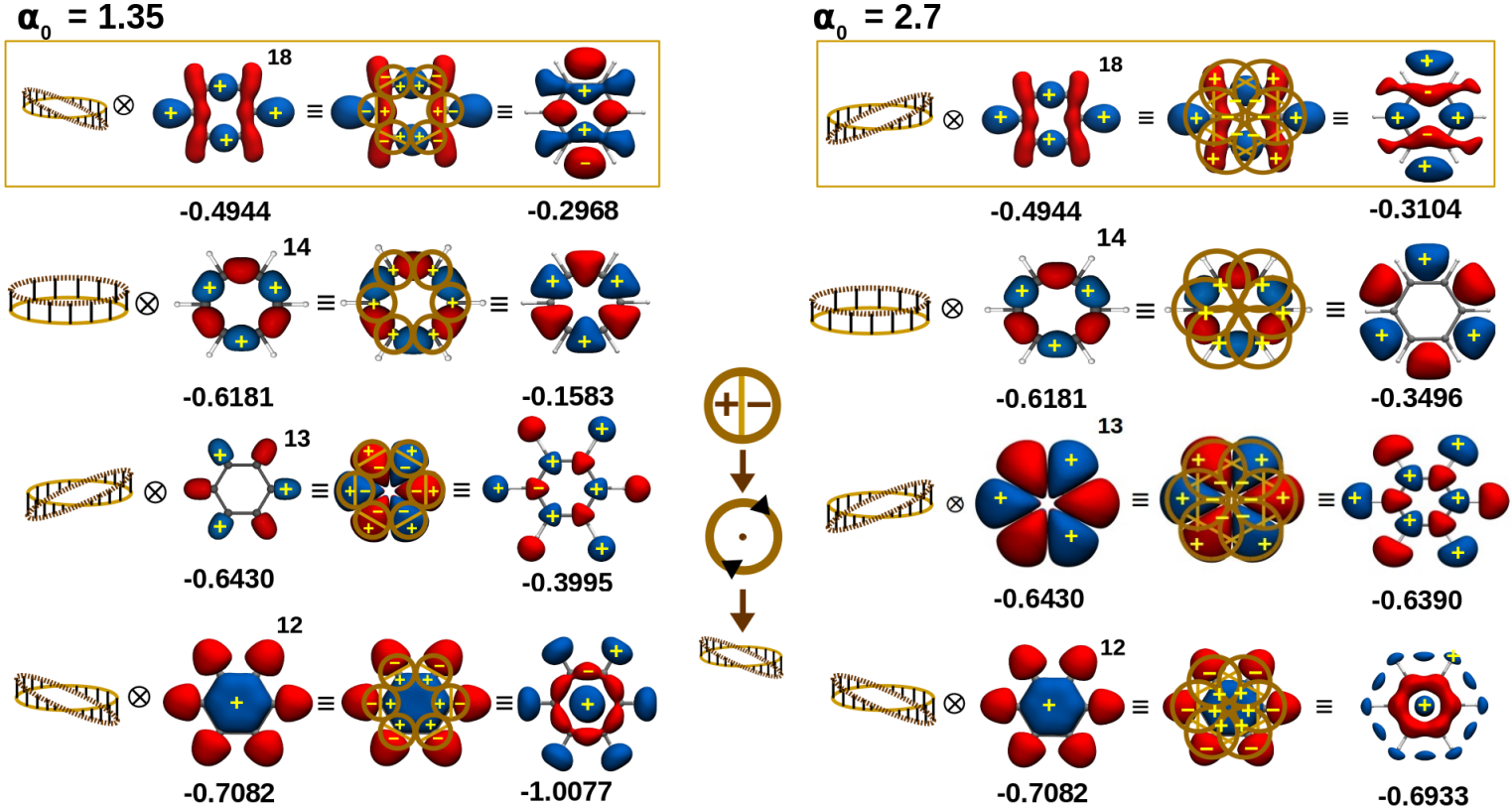}
\includegraphics[width=0.8\linewidth]{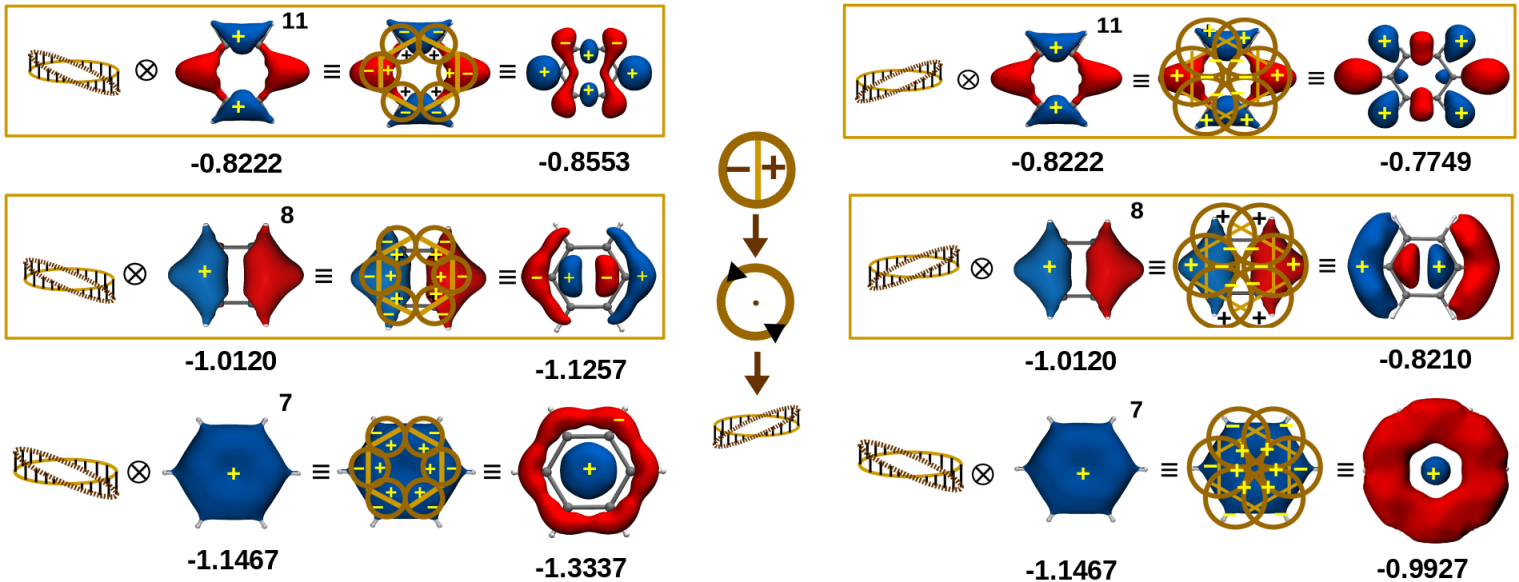}
\caption{Particle-in-a-ring (PIAR) analysis for MOs of laser-dressed benzene.  One sees a flip in the constituent PIAR states for MOs 8, 11 and 18, from clockwise at $\alpha_{0}$ = 1.35 a.u. to \textit{anti}-clockwise at $\alpha_{0}$ = 2.7 a.u. The ground state of a PIAR is totally symmetric while the higher excited states are doubly-degenerate. The first excited state is a degenerate single-node state, imposing an angular momentum in clockwise or \textit{anti}-clockwise directions on the field-free state upon switching on the laser.}
\label{fig:fig6}
\end{figure}

 Yet another evidence for the middle-of-the-ring localization is given by the Molecular orbitals in a mean-field picture of solving the KH time-independent Schr\"odinger equation. The MO's and MO energies as function of the parametric variation of $\alpha_{0}$, the laser parameter give reasons for the middle-of-the-ring electron localization. The mixing of the orbitals, due to the increasing $\alpha_{0}$ is evident in Figure~\ref{fig:fig5}. A Walsh-kind of diagram as a function of $\alpha_{0}$, instead of the nuclear geometry, is portrayed in Figures~\ref{fig:fig5}\textbf{(c)} and \textbf{(d)}. What is clearly evident are the following:
\begin{enumerate}
\item[\textbf{(i)}] The first six MO's having a total of 12 electrons are linear combinations of the $1s$-state on each carbon atom. For the field-free benzene, the nodal structure of these emanate from the interaction due to the nuclear geometry. In the case of the laser-dressed benzene, for $\alpha_{0}=1.35,1.50,2.7$ a.u. the circular electric field mixes these $s$-states, while preserving the nodal structure from the benzene geometry. For the ground state MO at $\alpha_{0}=1.35$ a.u., clearly the first MO density is being pushed into the ring and outwards, giving a circular dichotomy. The nodeless first MO at $\alpha_{0}$ is clearly an $s$-state with maximum density in the middle of the ring, evidencing a pair of electrons sitting at the ring-center. This state is clearly due to all the six tori centered on each carbon, intersecting to form a deep vortex in the middle which supports this state.
\item[\textbf{(ii)}] The next feature of interest are the $\pi$-orbitals of benzene, accounting for the 6 $\pi$-electrons, which are almost unaffected by the circular-polarization. This is evidently because $p_{z}$ orbitals have minimum density in the plane of the ring which coincides with the laser polarization. The aromaticity of benzene is therefore preserved in the presence of the CPL. The nodal structure from benzene is preserved for these orbitals.
\item[\textbf{(iii)}] The third feature is the distorted $\sigma$-framework of benzene, which is induced by the CPL, resulting in a $\sigma$-aromaticity on top of the unaffected $\pi$-aromaticity, giving a totally doubly aromatic character. These orbitals account for 30 electrons which follow the $4n+2$ count, where $n=7$. Here the interpretation of the rule is to be modified, with six contributions from the carbon atoms and an extra contribution from the $\sigma$- electrons displaced by the circularly polarized light which amounts to a $n=7$ count. These distorted $\sigma$-orbitals have an extra feature of CPL-induced circular nodal structure together with the nodal planes stemming from the benzene structure. The reason for this feature is explained in Figure~\ref{fig:fig6}, and it comes as a consequence of the circular dichotomy introduced by the CPL. The orbitals and thus the electronic structure therefore constitute a real laser-dressed state.
\end{enumerate}
\par
\begin{figure}[t]
\includegraphics[width=0.9\linewidth]{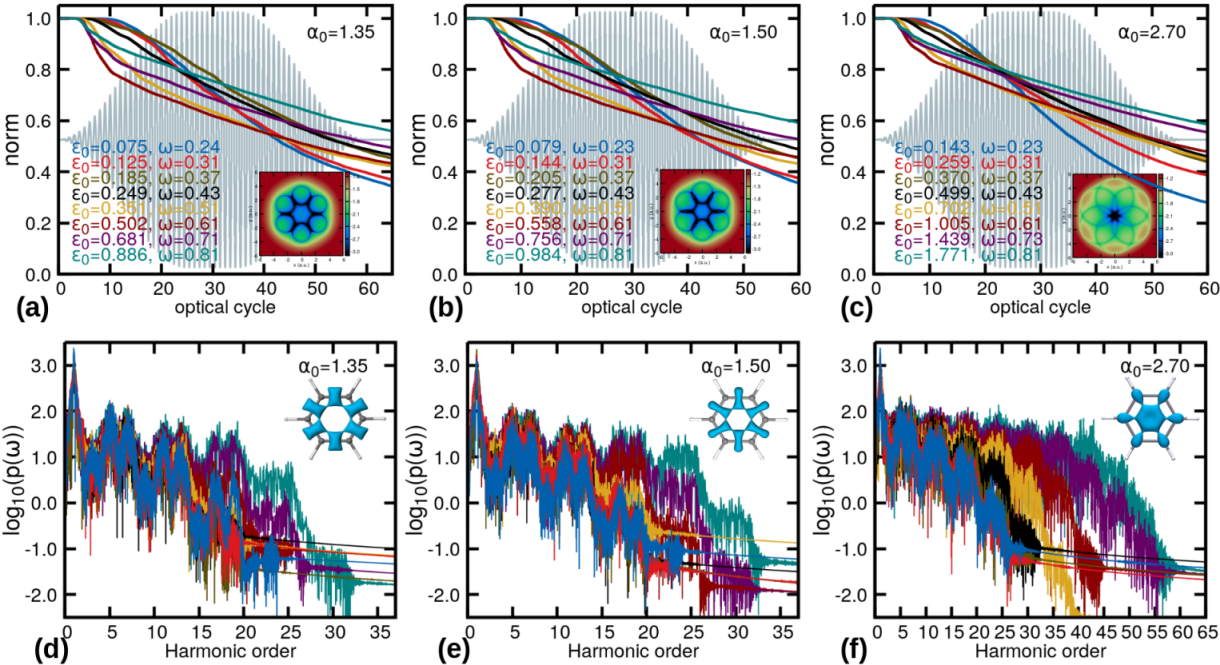}
\includegraphics[width=0.9\linewidth]{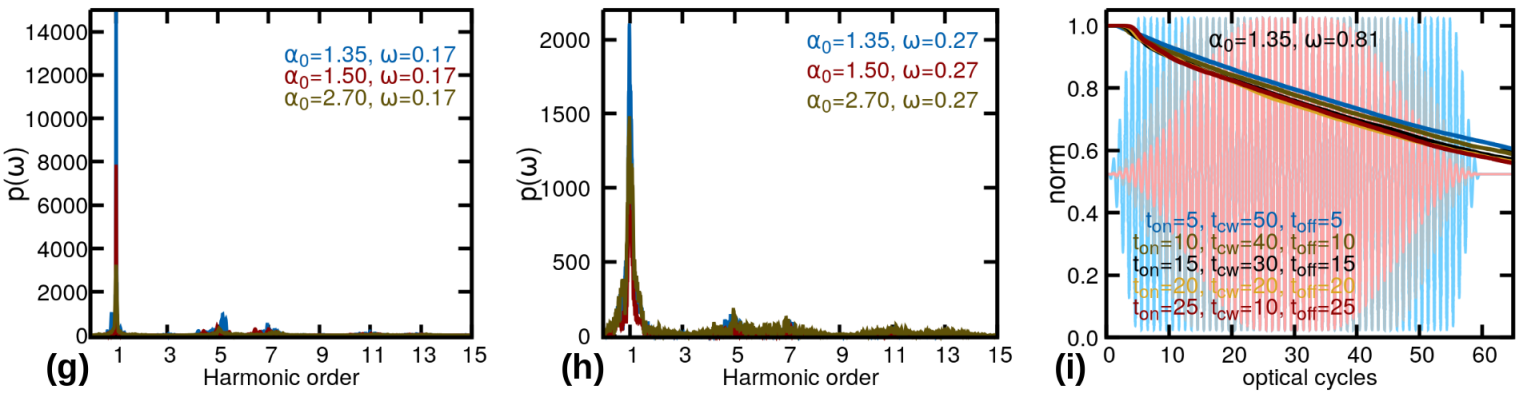}
\caption{(a)-(c) Norm of the wave-function with the number of optical cycles, for $\alpha_{0}$ 1.35, 1.50, 2.70. 
(d)-(f) Logarithm of the frequency-dependent intensity versus the harmonic order : Also, for $\epsilon_{0}$ = 1.771 a.u. and $\omega = 0.81$  a.u. at $\alpha_{0}$ = 2.7 a. u., the harmonic cut-off region extends  to  double of that  for $\alpha_{0}$ = 1.35 and 1.5 a.u. 
(h), (i) The most-intense peak is observed when $\alpha_{0} = 1.35$. Suggests a possible inference of C-C bond length in benzene from experimentally obtained spectra. 
(j) A short rise-time and a longer CW region in the pulse supports a greater conservation of the wave-function norm.
}
\label{fig:fig7}
\end{figure}

 Orbitals 1 through 6 register no change in the number of nodes in going from the field-free to with-field realm. However, MO 7 sees a gain of one node. The same is true for MOs 8-13 and MOs 15-18. MOs 14 and 19 do not gain a node upon switching on the laser. This nodal picture becomes lucid with a particle-in-a-ring analysis, presented in Figures~\ref{fig:fig7}. For instance, the field-dressed MO 7 can be visualized as a superimposition of a particle-in-a-ring first-excited state over a field-free MO 7.  This approach is extended to higher MOs and the constituent particle-in-a-ring states are depicted in the figures. The MOs which do not acquire a new nodal plane upon switching on the laser can be explained using a totally-symmetric particle-in-a-ring ground-state superimposed over the field-free MO. An intriguing feature is a flip in the direction of constituent ring states, from clockwise to counter-clockwise, for MOs 8, 11 and 18 when one goes from $\alpha_{0}$ = 1.35 a.u. to  $\alpha_{0}$ = 2.7 a.u.
\par The Hartree-Fock and MO energies are plotted for varying quiver distances and shown in Figure~\ref{fig:fig5}(\textbf{b}) and (\textbf{c}), respectively with an enlarged view of the latter in Figure ~\ref{fig:fig5}\textbf{(d)}. The core orbital (MO 1)  contributes to a  stabilized state owing to a visible dip in energy around $\alpha_{0}$ = 2.45 a.u. Even upon switching on the laser, the core MOs, 1-6, preserve their character and stay well below the higher valence MOs. A few of the low-lying and higher valence MOs dip and rise in energy around $\alpha_{0}$ = 1.35 a.u. However, these valence MOs return to more uniformly-spaced energy levels at $\alpha_{0}$ = 2.7 a.u., and contribute to a laser-dressed state of benzene which is bound. 
\par A simple way to understand the aforementioned nodal structure in the MOs forming the $\sigma$-framework is to think of the laser-dressed orbitals as a direct product of particle-on-a-ring states together with orbitals on benzene. This depiction is given in Figure~\ref{fig:fig6} for the MOs forming the $\sigma$ framework. It is as if there is a particle on a ring associated with each of the carbon atoms. When the rings start interacting with increasing $\alpha_{0}$, there are flips in the signs according to the mixing and the energetic stabilization. The degenerate excited states for particle-on-a-ring can be thought of as a classical \emph{clockwise} and \emph{anti-clockwise} circulation. This amounts to a local angular momentum imparted by the CPL. However, at $\alpha_{0}=2.7$, the net local angular momentum at the benzene ring-center is zero because of a localized $s$-state. The local angular momentum at each carbon atom is non-zero in some cases but the total vector addition will sum upto zero.
\par With all these exotic features including a middle-of-the-ring electron pair and double aromaticity, it is important to understand how these states can be realized in an experiment and what are their implications and applications. The nature of proton conduction when the benzene ring-center is electron-rich has already been discussed and this has implications for control over electrophilic substitution reactions. The double aromaticity also plays an important role here. However, a direct application is seen in terms of the harmonic generation spectra of these states for a given set of laser parameters. Firstly, we discuss an estimate of the laser parameters for intensity, frequency, risetime and pulse width of the circular polarized laser required for the formation of these states. The KH-Schr\"odinger equation in the oscillating frame of reference only gives the $\alpha_{0}$ parameter (see Equation~\ref{eq:eq06}), which will be a measure of the intensity for a fixed frequency.  To simulate experimental conditions, the solutions of a TDSE yielding the electronic dynamics in the lab-frame are presented in Figure~\ref{fig:fig7}, for $\alpha_{0}=1.35,1.5$ and $2.7$ a.u.  The intensity and frequency have been varied for a fixed $\alpha_{0}$. It is seen that for high values of $\epsilon_{0}=0.886$ a.u. (2.754 $\times$ 10$^{16}$ W cm$^{-2}$) and $\omega=0.81$ a.u. (22.04 eV), for $\alpha_{0}=1.35$ a.u., about 60\% of the benzene molecules survive the laser pulse and remain in this state as long as the peak CW region of the pulse is on. This is evident from the norm of the ground state wave function for all the cases.
\par Yet another factor that is analyzed is how fast should the rise time of the pulse be, so that these states have minimal ionization. This is portrayed in Figure~\ref{fig:fig7}\textbf{i}. The second panel in Figure~\ref{fig:fig7} gives the harmonic generation spectra due to the formation of these states. It is seen that the harmonic generation spectra are highly intense and with a large cut-off going upto 30 in the case of $\alpha_{0}=1.35$ and $\alpha_{0}=1.5$ a.u. Electronic structure-wise, both of these belong to category (ii) discussed at the beginning of this section. What seems striking is that when one has a middle-of-the ring localization for $\alpha_{0}=2.7$ a.u. the cut-off in the harmonic order dramatically increases to 55 as seen in Figure~\ref{fig:fig7}\textbf{(f)}, where the intensity and frequency are such that the long lived state is formed. This is evidently due to the new bound state of the paired up electrons in the middle of the ring which are bound by the CPL.
\section*{Concluding remarks}
\par Our work has discussed the formation of exotic states for benzene in the presence of circularly polarized light, with plane of light polarization coinciding with ring plane in benzene. Through electronic dynamics at a fixed nuclear geometry, we have shown that such states can be experimentally obtained using a high intensity XUV-laser. Under high frequency conditions, as high as 16-22 eV at least 60\% of the benzene molecules survive the laser pulse. The laser modifies the electronic structure of benzene such that the exotic states formed through a laser-induced hybridization have very interesting properties which include \textit{two} electrons sitting in the middle of the benzene ring and a double aromaticity. 
\par The laser parameters obtained fall in line with current experimental capabilities. Previous works had shown ionization and Coulomb explosion of benzene when the intensity was of the order of $10^{16}$ W/cm$^{2}$ at much lower frequencies of 1 eV to 8 eV. The presence of the high frequency stabilization given in our work, enables us to put forth a chemically modified benzene through laser-induced hybridization of its orbitals, that has properties very different from that of the usual benzene. It has been shown here through numerical calculations of properties and energetics that the laser-dressed benzene shown can stabilize a bare proton in the middle of the ring. With a benzene-proton complex forming a precursor for the Wheland ion intermediate, the new state predicted in the presence of the laser can imply temporal attosecond control over the dynamics of electrophilic substitution reactions. 
\par Of late, improbable reactions being catalyzed by very strong static electric fields have been a subject of interest. A strong DC field is difficult to produce in a laboratory, however a time-dependent attosecond pulse constituting an oscillating electric field is nowadays state of the art. Moreover, the oscillating electric field has control over the ionization dynamics as well.
\par The experimental manifestation and effects of the xotic state have direct applications in terms of higher harmonics produced. The laser-dressed high frequency state in the case of benzene, with minimal ionization can produce a very large number of harmonics going upto 55 in comparison with the low intensity ionizing state where the harmonics last upto about 30. This again is due to the exotic set of bound states formed in the high-frequency stabilization.
\par Very recently light-induced charge density waves have been experimentally reported in a layered solid, LaTe$_{3}$. Again graphene and its laser-dressed states have had a recent experimental manifestation. A natural extension of our work related to benzene is to apply the same for graphene. In the case of graphene too, with circularly polarized light in the plane of the layer, instead of the light breaking the symmetries, \emph{time crystals}, with structural and temporally periodic properties would certainly form a novel state of matter, the light-dressed state.
\section*{Computational methodology} 
Time-independent zeroth-order KH calculations in the effective potential $V^{KH}_{0}$of benzene have been done using GAMESS-US\cite{Gordon1993} package with a COEMD-3 basis\cite{Susi2013} and an extra set of $d$-type even-tempered basis functions on the carbon atoms. Since, the KH-transformation only modifies the nuclear-electron attraction term, the GAMESS-US\cite{Gordon1993} has been modified for the KH calculations. The integration for the zeroth order terms has been done using Gauss-legendre method with 501 number of grid points. Time-independent calculations have been done using 534 Cartesian basis functions at the Hartree-Fock (HF) level of theory, which is a sufficient description for the mean-field orbitals and the density. 
\par Time-dependent Schr\"odinger equation for the many-electron system under the dipole approximation  in the lab frame and length gauge will be:
 \begin{eqnarray}
i\frac{\partial}{\partial t}\Psi\left(\{\vec{r}_{j}\},t\right)=\left[ -\frac{1}{2}\sum_{j}^{n}\nabla^{2}_{j} - \sum_{A}^{M}\sum_{j}^{n}\dfrac{Z_{A}}{\left| \vec{r}_{j} - \vec{R}_{A} \right|} + \sum_{j>k}^{n}\frac{1}{\left|\vec{r}_{j}-\vec{r}_{k}\right|} - \sum_{j=1}^{n}\vec{r}_{j}.\vec{\epsilon}(t) \right]\Psi\left(\{\vec{r}_{j}\},t\right) \nonumber
\label{TDSE}
\end{eqnarray}
 This has been used for the electronic dynamics and the harmonic generation spectra presented in the work. Time-dependent calculations have been done with 534 Cartesian basis functions which describe 4201 configuration state functions (CSF) using $(t,t^{\prime})$-method. The MO-transformed Configuration interaction singles (CIS) Hamiltonian, dipole and complex absorbing potential (CAP) matrix have been extracted out from GAMESS-US. Our calculation therefore only includes singly excited states. The number of states taken is large and sufficient enough to describe the dynamics in the presence of the high frequencies that have been presented. The box type CAP matrix has been used for the ionization.\cite{Santra2001, Krause2014, Schlegel2014} The CAP has been put outside the classical turning points at $x=19$ a.u., $y=19$ a.u. and $z=13$ a.u. The CAP integrals have been checked to be non-vanishing. The $(t,t^{\prime})$ method was developed by Peskin and Moiseyev.\cite{Peskin1993, Peskin1994, Peskin19942, Prashant2019} Here, a new fictitious time co-ordinate, denoted as $t^{\prime}$, is introduced in an extended Hilbert Space as $i\hbar\left [\frac{\partial}{\partial t}+\frac{\partial}{\partial t'}\right]\Tilde{\Psi}(\vec{r},t,t^{\prime})=\hat{H}(\vec{r},t')\Tilde{\Psi}(\vec{r},t,t^{\prime})$. On the contour $t=t^{\prime}$, this equation becomes equal to the physical time-dependent Schr\"odinger equation. It can be rearranged in the form:
\begin{equation}\label{tt'derive}
i\frac{\partial}{\partial t}\Tilde{\Psi}(\vec{r},t,t^{\prime})=\hat{{\cal H}}_{\cal F}(\vec{r},t^{\prime}) \Tilde{\Psi}(\vec{r},t,t');~\hat{{\cal H}}_{\cal F}(\vec{r},t^{\prime})=\left[\hat{H}(\vec{r},t^{\prime})-i\hbar\frac{\partial}{\partial t'}\right ] \nonumber
\end{equation} 
Here, a $\hat{{\cal H}}_{\cal F}$ resembles a Floquet type operator in the $t^{\prime}$ coordinate. The evolution equation of the wave function now takes the form:
\begin{equation}\label{tt'evolution}
\Tilde{\Psi}(\vec{r},t,t^{\prime})=e^{-{i}\hat{{\cal H}}_{\cal F}(\vec{r},t^{\prime})(t-t_0)}\Tilde{\Psi}(\vec{r},t_0,t^{\prime}). \nonumber
\end{equation}
The physical solution $\Psi(\vec{r},t)$ can be extracted from the full solution $\Tilde{\Psi}(\vec{r},t,t^{\prime})$ at $t=t^{\prime}$ and all the physical properties can then be calculated. The time-dependent Hamiltonian has been solved for 65-optical cycles with a time step, $dt$ of 0.01 a.u. High harmonic generation (HHG) spectra have been calculated using time-dependent induced dipole moments. The Fourier transform of the double derivative of the induced dipole moment gives the Harmonic generation spectra.

\begin{acknowledgement}
\par P. R. thanks the INSPIRE program for a research fellowship. M. P. acknowledges IISER-M for a research fellowship. M. R., currently affiliated with  University of Southern California, Department of Chemistry, Los Angeles, California,  acknowledges the INSPIRE program for an erstwhile fellowship at IISER-M.  P. B. thanks SERB, Department of Science and Technology (DST), for a grant through a project entitled \emph{Chemical Reactions in Oscillating Fields}. The authors acknowledge the High-Performance Computing (HPC) facility at IISER-M. PB thanks  Professor Lorenz Cederbaum(Universitat Heidelberg) for coming up with the question: \emph{what are the interesting KH states of atoms and molecules in a circularly polarized laser?}. Discussions with Professor Srihari Keshavamurthy at Indian Institute of Technology(IIT)-Kanpur are gratefully acknowledged. Taseng Mancheykhun and Alkit Gugalia (QM group, IISER-M) were instrumental in preparing the TOC and the tagline.
\end{acknowledgement}

\bibliography{achemso-demo}

\end{document}